\documentclass[%
 reprint,
 amsmath,amssymb,
 aps
]{revtex4-2}

\usepackage{amsmath}
\usepackage{booktabs}
\usepackage{multirow}
\usepackage{color}
\usepackage{xr}
\usepackage{graphicx}% Include figure files
\usepackage{dcolumn}% Align table columns on decimal point
\usepackage{bm}% bold math
\begin{document}

\preprint{APS/123-QED}

\title{Diret observation of strong t-e orbital hybridization and the effects of f orbitals }% Force line breaks with \\%

\author{Mian Wang}
\author{Qian Zhang}
\author{Shuai Jing}
\author{Xiang-Guo Li}
 \email{lixguo@mail.sysu.edu.cn}
\author{Yanglong Hou}
\affiliation{%
School of Materials, Sun Yat-sen University, Shenzhen 518107, China 
}%

\date{\today}% It is always \today, today,
             %  but any date may be explicitly specified

\begin{abstract}
Recent research has revealed that the Cr family perovskite ReCrO$_3$ exhibits intriguing magnetic coupling interactions within Cr pairs, which may not follow the Goodenough-Kanamori (GK) rules due to the t-e hybridization between Cr$^\mathrm{III}$ ions. We investigate the complex magnetism involving both t-e hybridization and Re-$f$ orbitals in the molecular analogue of perovskite  [$\mathrm{Ce_2^{III}Ce^{IV}Cr_8^{III}O_8(O_2CPh)_{18}(HO_2CPh)}$] ($\mathrm{Ce_3Cr_8}$) using first-principles method. Our results have shown that distinct from the bulk ReCrO$_3$, the superexchange via Cr-$d$ and O-$p$ orbitals can exhibit a unexpected dominate ferromagnetic (FM) Cr-O-Cr superexchange interaction in $\mathrm{Ce_3Cr_8}$ due to the strong t-e hybridization originated from the distorted molecular structure. The great sensitivity of the t-e hybridization with respect to the molecular structure, e.g., the angle of Cr-O-Cr, can lead to a ground state transition from ferromagnetic to antiferromagnetic state with the changes in the angle of Cr-O-Cr. The Ce-$f$ orbitals near the Fermi level can reduce this sensitivity through interacting with the Cr-$d$ orbitals via the virtual charge transfer process. Our results are strongly supported by a modified superexchange model based on the t-e hybridization theory. These findings complete the theory of superexchange magnetism involving the t-e hybridization and $f$ orbitals, and in the meanwhile introduce a new avenue for fine-tuning the magnetic characteristics via Tm-d/Re-f interactions at nanoscale.
\end{abstract}

%\keywords{Suggested keywords}%Use showkeys class option if keyword
                              %display desired
\maketitle

%\tableofcontents

%\section{\label{sec:Introduction}Introduction}
Magnetic perovskite materials continue to attract widespread attentions in the scientific community due to its excellent and fascinating physical properties such as colossal magnetoresistance and multiferroicity \cite{narayan2019multiferroic,wang2023towards,ricciardulli2021emerging,spaldin2019advances,wang2015observation,li2017chemically,schmid1994multi}. In particular, the chromium-based RECrO$_3$ (RE = rare earths)compounds show a broad application in the fields of catalyst, thermistor, fuel cell, and nonvolatile memory devices because of their intriguing magnetic and ferroelectric properties\cite{zhu2022crystal}. However, understanding the underlying magnetic exchange mechanisms still present great challenges due to its complexity. For example, a ferromagnetism (FM) from the t-e hybridization between Cr$^{\mathrm{III}}$ ions was reported previously in RECrO$_3$ \cite{zhou2010intrinsic,zhou2011magnetic,moon2017structural,zhu2022crystal}, while the evidence is indirect and not adequate since the antiferromagnetic (AF) superexchange interaction is still the main contribution in the exchange path of Cr$^{\mathrm{III}}$-O-Cr$^{\mathrm{III}}$ with bridging angle much greater than 90\textdegree~ in bulk RECrO$_3$, according to Goodenough-Kanamori rule\cite{goodenough1955theory,goodenough1958interpretation,kanamori1959superexchange}. The dominance of the FM superexchange interaction in such geometry, that is the direct evidence of the existence of t-e hybridization induced FM interaction, has never been achieved. Not to mention the quantitative analysis of this FM interaction.

The recently synthesized molecular analog of the perovskite repeating unit [Ce$_2^{\mathrm{III}}$Ce$^{\mathrm{IV}}$Mn$_8^{\mathrm{III}}$O$_8$(O$_2$CPh)$_{18}$(HO$_2$CPh), abbreviated as Ce$_3$Mn$_8$], which manifests rich and complex physics with a variety of magnetic interactions involving $f$ electrons\cite{thuijs2017molecular}, provides a new platform for the underlying exchange mechanism investigation. Introducing an extra $d_\sigma$ electron by substituting Mn$^{\mathrm{III}}$ with Fe$^{\mathrm{III}}$ can result in more pronounced asymmetric behavior in the FM interaction involving transition metal (TM) 3d and RE 4f orbitals \cite{wang2023molecular}. Cr$^{\mathrm{III}}$, on the other hand, without occupied $d_\sigma$ orbitals can potentially exhibit more novel physics. For example, the more distorted molecular framework compared to its bulk counterpart may present a different degree/form of t-e hybridization. The RE-f orbitals can also potentially have significant effects on the t-e hybridization. Such physics have never been realized and investigated yet. In addition, the nanoscale size of the molecule enables the fine-tuning of its properties by chemical doping or external manipulation, which is intrinsically different from its corresponding bulk perovskite. Consequently, it is desirable to study these fundamental physics related to the magnetic exchange mechanisms in the molecular form of RE-Cr systems.

In this letter,  we theoretically explore the complex magnetism in Ce$_3$Cr$_8$, a molecular analogue of the perovskite repeating units by replacing Mn ions with Cr ions in Ce$_3$Mn$_8$ \cite{thuijs2017molecular,wang2017cation}, with main focus on the t-e hybridization and the effects of $f$ orbitals using the first-principles method. Interestingly, our calculations show that FM exchange coupling between Cr ions via t-e hybridization is the dominant interaction in Ce$_3$Cr$_8$ leading to a FM ground state, distinct from its bulk RECrO$_3$ crystal in which AF interaction is still the leading interaction. The calculated exchange strengths at various Coulomb energies and Cr-O-Cr bond angles quantitatively agree with a simple model derived from orbital overlap integrations, further confirming the FM nature from t-e hybridization. More importantly, the Ce$^{\mathrm{IV}}$-f orbital around the fermi level can significantly change the sensitivity of strain induced exchange coupling changes through the virtual charge transfer between Ce-4f orbitals and Cr-3d orbitals, different from the direct FM contribution from f orbitals in Ce$_3$Mn$_8$ \cite{thuijs2017molecular} and Ce$_3$Fe$_8$ \cite{wang2023molecular} in our previous investigations. Our results add extra content for the fundamental superexchange coupling physics by directly demonstrating the dominant FM interaction in TM-O-TM with bonding angle much greater than 90\textdegree. The effects of f orbitals on such exchange mechanism can provide another dimension to modify and control the magnetic behavior in magnetic molecular-based devices.

\begin{figure*}[t]
\centering
\includegraphics[width=0.9\linewidth]{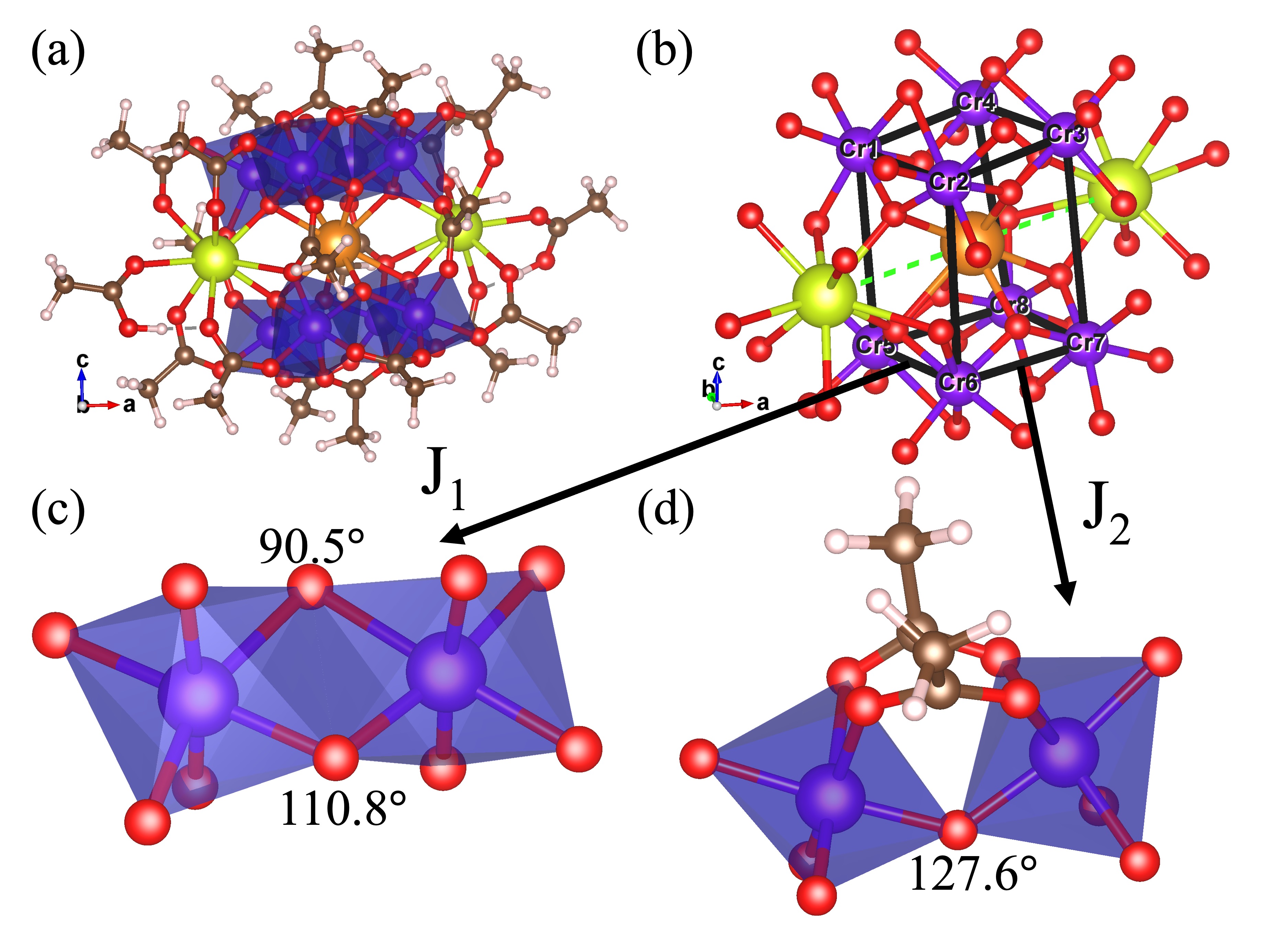}
\caption{The optimized molecular structure of Ce$_3$Cr$_8$. (a) The complete molecular structure of Ce$_3$Cr$_8$ with -CH$_3$ as the ligand group. The CrO$_6$ octahedra are shaded in blue. Color scheme: Ce$^{\mathrm{IV}}$ orange; Ce$^{\mathrm{III}}$ green; Cr$^{\mathrm{III}}$ purple; O red. 
(b) The partially Ce$_3$Cr$_8$ central fragment without ligands, showing only the Ce-O and Cr-O bonds. The Ce line is denoted by a dotted line. The Cr curb is labelled by black line. (c) The detailed structure of J$_1$ (d)The detailed structure of J$_2$, The average degrees of $\angle$Cr-O-Cr are labeled in fig.}
\label{fig_structure}
\end{figure*}

%\section{\label{sec:Method}Method}
\textit{Method.--} Our first-principles calculations are carried out within the framework of Kohn-Sham Density Functional Theory(DFT)\cite{kohn1965self} with the generalized gradient corrected Perdew-Burke-Ernzerhof (PBE) exchange-correlation functiongal \cite{perdew1996generalized} using the Vienna Ab-initio Simulation Package(VASP) code \cite{kresse1996efficiency,kresse1996efficient}. The electron-ion interaction was described using projector augmented wave(PAW) potentials\cite{blochl1994projector,kresse1999ultrasoft}. The energy cutoff for plane-wave basis expansion was set to 500 eV. The threshold for self-consistency and structure optimization were set to $10^{-5}$~eV and 0.01~eV/\AA, respectively. The molecule was put into a large periodic supercell of 28~\AA$\times$22~\AA$\times$24~\AA~to make sure the distance between the molecule and its repeating image is larger than 10~\AA. Because of the strong localization of Ce-$f$ and Cr-$d$ orbitals, the GGA+U method was applied with U$=$2.0~eV\cite{loschen2007first,thuijs2017molecular,wang2023molecular} and 3.0~eV\cite{hong2012spin} for the Ce-$f$ and Cr-$d$ orbitals, respectively. The Wannier90 package \cite{mostofi2008wannier90} was applied to calculate Wannier Functions. The spin-orbital coupling is switched off due to its negligible effects on the exchange coupling in 3d- and 4f-compounds\cite{zhang2018stabilization,shorikov2005magnetic,wang2023molecular}.

\iffalse
\begin{figure*}[hbt]
\centering
\includegraphics[width=0.9\linewidth]{fig_mag.png}
\caption{(a) Eight spin-ordering configurations in a Ce$_3$Cr$^{\mathrm{III}}_8$ molecule,derived from four known configurations in a perovskite unit cell(b)The multi-spin Heisenberg model showing the magnetic exchange coupling paths labeled as $J_1$ to $J_4$; lines in the same color indicate symmetry-equivalent paths.  Other possible paths are unlabeled because of the exchange coupling strengths at least an order of magnitude smaller than the ones labeled in the figure. }
\label{fig_mag}
\end{figure*}
\fi

{\textit{Results.--} The initial structure of Ce$_3$Cr$_8$ molecule is obtained by substituting Cr ions for Mn ions in the recently reported Ce$_3$Mn$_8$ molecule \cite{thuijs2017molecular}. Analogy to Ce$_3$Mn$^{\mathrm{III}}_8$, the Ce$_3$Cr$^{\mathrm{III}}_8$ molecule have a striking structural similarity to the repeating unit of perovskite, which resembles a repeating unit of ABO$_3$ cubic with distortions and plus two A ions, as shown in Figure~\ref{fig_structure}(a) and (b). The core of Ce$_3$Cr$^{\mathrm{III}}_8$ includes eight Cr sites and three Ce sites. Three Ce ions are arranged in a line as shown in Fig.~\ref{fig_structure}b. The central Ce ion has an oxidation state of +4 and is eight-coordinated, while the other two Ce ions have an oxidation state of +3 and are nine-coordinated. The eight Cr$^{\mathrm{III}}$ ions in Ce$_3$Cr$^{\mathrm{III}}_8$ can be divided into two groups, each group has four Cr$^{\mathrm{III}}$ ions, which are separated by the Ce line, denoted as the top group(Cr1, Cr2, Cr3, Cr4) and the bottom group (Cr5, Cr6, Cr7, Cr8). 

We perform the subsequent electronic structure calculations after a fully geometry optimization. We mainly focus on the nearest-neighbor(NN) exchange couplings of the Cr ions along the paths involving Cr-O-Cr geometries, including J$_1$ perpendicular to the Ce line (see Fig~\ref{fig_structure}c) and J$_2$ parallel with the Ce line (see Fig~\ref{fig_structure}d). In the neighboring CrO$_6$ of J$_1$, the two Cr-O octahedras are edge-sharing with two bridging oxygen atoms with Cr-O-Cr bonding angles 110.8\textdegree and 90.5\textdegree, respectively(see Fig~\ref{fig_structure}c); the neighboring CrO$_6$ within the J$_2$ share one corner with one bridging oxygen atom with Cr-O-Cr bonding angles 127.6\textdegree, and two carboxylate groups connect the two Cr ions within the neighboring CrO$_6$ octahedras(see Fig~\ref{fig_structure}d).

%\subsection{\label{sec:Magnetic}Magnetic structure and first-principles energetics.}

The perovkite have four known common type of spin-ordering configurations: FM, A-AF,C-AF and G-AF \cite{shein2005band}. While Ce$_3$Cr$_8$ molecule has low symmetry with three inequivalent interaction planes or axes along different orientations, leading to three distinct A-AF and three different C-AF states. Thus there are eight high symmetrical magnetic configurations (plus one G-AF and one FM) in Ce$_3$Cr$_8$, illustrated in the supplementary Figure~S1. Our results show that FM is the ground state of Ce$_3$Cr$_8$ molecule by performing DFT calculations for all eight magnetic configurations. The calculated total energies of these eight magnetic configurations are listed in Supplementary Table~S1 . The eight Cr ions in Ce$_3$Cr$_8$ have 2.8 $\mu_B$ magnetic moment for each Cr which indicate an oxidation state of $+3$ \cite{abbad2016search}. The two Ce$^{\mathrm{III}}$ ions outside the Cr curb have a magnetic moments of 1~$\mu_B$ for each, and the central Ce$^{\mathrm{IV}}$ has a small magnetic moments (less than 0.1~$\mu_B$). 

To reveal the underlying magnetic properties of Ce$_3$Cr$_8$, the Cr/Cr exchange coupling parameters ($J$) are estimated by a multi-spin Heisenberg model \cite{noodleman1981valence,noodleman1986ligand,noodleman1995orbital,yamaguchi1988ab,yamaguchi1989antiferromagnetic} with total energies of different magnetic states. The spin Hamiltonian is defined as, 

\begin{equation}
\begin{aligned}\label{eq:first}
\hat{H}=-\sum_{i<j}J_{ij}\vec{S_i}\cdot\Vec{S_j}
\end{aligned}
\end{equation}
where $J_{ij}$ are magnetic coupling parameters between the Cr ions at sites $i$ and $j$; $\vec{s_i}$ and $\vec{s_j}$ are the spin vectors of Cr ions at site $i$ and $j$ , respectively. The total spin of S is $1.5$ for each Cr ion according to the Hund’s rule.  Our total energy results indicate that two spin coupling paths, $J_1$ to $J_2$ (see Fig.~\ref{fig_structure}), exhibit significant contributions to the total energy. Other coupling paths, e.g., the diagonal direction of the side faces denoted as $J_3$ and $J_4$, are shown in Supplementary Materials Fig.~S1, which are much smaller compared to $J_1$ and $J_2$ (see Supplementary Material section I). The calculated coupling strengths (see Table \ref{table_J}) are $J_1=+1.09$~meV and $J_2=+2.07$~meV (positive for FM, negative for AF). The strong FM interactions of $J_1$ and $J_2$ result in the FM ground state. 

The dominant FM nature of $J_1$ and $J_2$ is unexpected after analysing the superexchange mechanisms. For example, in $J_2$ exchange coupling pathway, the superexchange through the Cr-O-Cr($\angle$Cr-O-Cr=127.6) should contribute AF interactions according to GK rules \cite{thuijs2017molecular,zhu2022crystal}. The reported FM contribution from the the t$_{2g}$-e$_g$(t-e) hybridization on the Cr$^{\mathrm{III}}$-O-Cr$^{\mathrm{III}}$ couplings via the virtual charge transfer (VCT) of t$_{2g}^3$-O-e$_g^0$ is much weaker compared to the AF interaction in bulk RECrO$_3$ crystal \cite{zhou2010intrinsic}. As such, $J_2$ should exhibit a AF nature, or at least not FM-dominated nature. We attribute this strong FM nature to the more distorted molecular structure compared to its bulk crystal, which can induce much larger tilting in the Cr-O octahedron and lead to much stronger t-e hybridization. This assessment is further confirmed by quantitatively evaluating the competing mechanisms between AF and FM in $J_2$, as shown next.

In the framework of t-e hybridization, the superexchange interactions of Cr$\mathrm{^{III}}$ ions include two parts: the superexchange interactions over the half-filled $\pi$ bond t$_{2g}^3$-O-t$_{2g}^3$, which is AF according to GK rules; the interaction between the half-filled $\pi$ bond and empty $\sigma$ bond, i.e., t$_{2g}^3$-O-e$_{g}^0$, due to the introduction of a VCT to the empty $\sigma$ bond from t-e hybridization, which shows the FM coupling (see Supplementary materials section II and Fig.~S2 for detailed information and the corresponding schematic hybridization diagram). The sum of this two parts gives the total interaction $J$ in the $J_2$ path, as shown below\cite{zhou2010intrinsic},
\begin{equation} 
\label{eq:second} 
\begin{aligned}
J=J^\sigma-J^\pi
\end{aligned} 
\end{equation}
where J$^\pi$ expresses the AF interaction via the VCT of t$_{2g}^3$-O-t$_{2g}^3$ and the J$^\sigma_{hb}$ denotes FM interaction via the VCT of t$_{2g}^3$-O-e$_{g}^0$ due to t-e hybridization. They can be further expressed in an analytical form by considering all possible orbital overlap integrations between O and Cr \cite{zhou2010intrinsic} (also see Supplementary section II for detailed derivation), as shown below,
\begin{equation} 
\label{eq:third} 
\begin{aligned} 
J &= J_0[ \eta(c^\sigma)^2-(c^\pi)^2]
\end{aligned} 
\end{equation}
\begin{equation} 
\label{eq:forth} 
\begin{aligned} 
J_0 &= \frac{(V_{pd}^\pi)^2}{U + \triangle_{ex}}\\
\end{aligned} 
\end{equation}
\begin{equation} 
\label{eq:fifth} 
\begin{aligned} 
\eta &= \frac{(V_{pd}^\sigma)^2}{(V_{pd}^\pi)^2} \cdot \frac{U + \triangle_{ex}}{U + \triangle_{c}} \\
\end{aligned} 
\end{equation}
\begin{equation}
\begin{aligned}
c^\sigma = & [ \cos \left( \frac{w}{2} \right) + \sin \left( \frac{w}{2} \right)] \{ \sqrt{3} \cos \left( \frac{w}{2} \right) \sin \left( \frac{w}{2} \right) +\\
&\begin{split}
\frac{\sqrt{3}}{2} [ \cos^2 \left( \frac{w}{2} \right) - \sin^2 \left( \frac{w}{2} \right) ] - \frac{1}{2} \}
\end{split}
\end{aligned}
\label{eq:sixth}
\end{equation}

\begin{figure*}[t]
\centering
\includegraphics[width=0.9\linewidth]{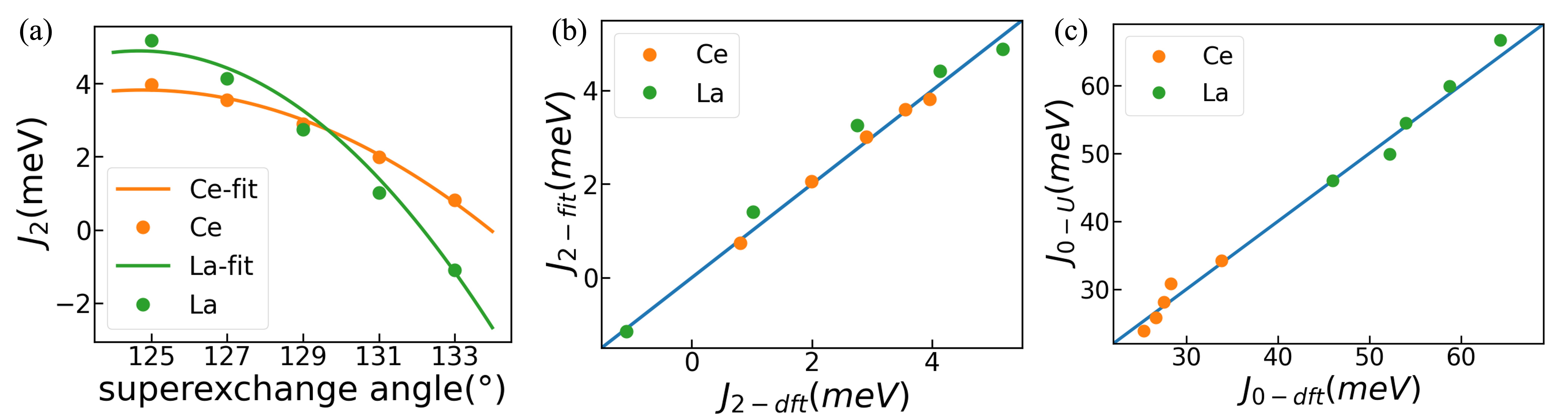}
\caption{Plot of superexchange strength from DFT calculated and model fitted values in Ce$_3$Cr$_8$ and [La$_3$Cr$_8$]$^{-1}$. (a) $J_2$ from DFT versus the superexchange angle Cr-O-Cr; (b) $J_2$ from model fitted values according to Eq.~\ref{eq:third} ($J_2-fit$) versus DFT calculated values ($J_2-dft$); (c)$J_0$ calculated from Eq.~\ref{eq:forth} ($J_0-U$) versus fitted from DFT results according to Eq.~\ref{eq:third} ($J_0-dft$). }
\label{fig_fit}
\end{figure*}

\begin{equation} 
\label{eq:seventh} 
\begin{aligned} 
c^\pi &= 2 \left[ \sin \left( \frac{w}{2} \right) + \cos \left( \frac{w}{2} \right) \right] - 4 \sin \left( \frac{w}{2} \right) \cos^2 \left( \frac{w}{2} \right) \\
\end{aligned} 
\end{equation}
where J$_0$ is a prefactor depending on orbital overlap integral V$_{pd}^\pi$, the on-site Coulomb $U$ and the exchange splitting $\triangle_{ex}$; $\eta$ is a dimentionless parameter defined as the multiplication of two ratios, an orbital overlap integral ration ($\frac{(V_{pd}^\sigma)^2}{(V_{pd}^\pi)^2}$) and an energy ratio ($\frac{U + \triangle_{ex}}{U + \triangle_{c}}$). $\triangle_{c}$ is the crystal-field splitting; c$^\pi$ and c$^\sigma$ are parameters evolving with the bridging angle of $\angle$Cr-O-Cr ($w$) \cite{zhou2010intrinsic,zhu2022crystal,slater1954simplified}. In order to quantitatively determine the origin of FM in Ce$_3$Cr$_8$, we calculated a series of $J_2$ values (see Table \ref{table_J}) by computing the total energies of the eight magnetic states at different cases (see Supplementary materials Table.~S2 for the detailed results at each case), including 1) Ce$_3$Cr$_8$ molecule with different angles of $\angle$Cr-O-Cr in $J_2$ (125\textdegree, 127\textdegree, 129\textdegree, 131\textdegree, 133\textdegree); 2)[La$_3$Cr$_8$]$^{-1}$ molecule with different angles of $\angle$Cr-O-Cr in $J_2$ (125\textdegree, 127\textdegree, 129\textdegree, 131\textdegree, 133\textdegree). The [La$_3$Cr$_8$]$^{-1}$ molecule is obtained by replacing the there Ce ions with La ions and adding an additional electron to keep the valence state of Cr ions unchanged. The chosen of [La$_3$Cr$_8$]$^{-1}$ is stimulated by our previous work\cite{thuijs2017molecular} showing that the unoccupied Ce$^{\mathrm{IV}}$-$f$ orbital can enhance the FM interaction between transition metals after comparing the exchange coupling strengths between [Ce$_3$TM$_8$] and [La$_3$TM$_8$]$^{-1}$.

With the data of angle dependent $J_2$ in Table~\ref{table_J}, we then fit Eq~\ref{eq:third} by setting $J_2$ ($J$ in Eq.~\ref{eq:third}) as the y values, angle ($w$) as the x values, and $J_0$ and $\eta$ as the fitting parameters. An excellent fit was obtained by observing the fitted curves as well as the corresponding parity plot (see Fig.~\ref{fig_fit}a, b). This clearly indicates the validity of Eq.~\ref{eq:third} in describing the exchange mechanisms in Ce$_3$Cr$_8$ and La$_3$Cr$_8$. The fitted parameters for $J_0$ and $\eta$ are 26.63 meV, 1.86 and 52.22 meV, 1.81 for Ce$_3$Cr$_8$ and La$_3$Cr$_8$, respectively. The $J_0$ for Ce$_3$Cr$_8$ is about half of that in La$_3$Cr$_8$, while $\eta$ is almost unchanged, whose value is also reasonable\cite{zhou2010intrinsic}. The value of $J_0$ determines the sensitivity of the change of $J_2$ with respect to the superexchange angle TM-0-TM. With the increase of the TM-O-TM angle, the exchange coupling can eventually turn to AF dominated, e.g. the studied superexchange $J_2$ (see Table.~\ref{table_J}). Thus the exchange coupling in Ce$_3$Cr$_8$ does not always show more FM nature than that in La$_3$Cr$_8$, which is different from Ce$_3$Mn$_8$ \cite{thuijs2017molecular} or Ce$_3$Fe$_8$ \cite{wang2023molecular}. The validity of Eq. \ref{eq:forth} is further demonstrated with the following steps: 1) Selecting a series of $U$ values (1.5 eV, 2.0 eV, 2.5 eV, 3.0 eV, 3.5 eV) and at each $U$ value, calculate angle-dependent $J_2$ values (see Supplementary material Table~S3 and Table~S4 for the detailed calculation data) and fit Eq.~\ref{eq:third} to obtain the fitted $J_0$ for each $U$ value; 2) Estimate $\triangle_{ex}$ by constructing the maximally localized Wannier functions (MLWFs) based on the ferromagnetic electronic structure\cite{liu2016orbital}. Our estimated values of $\triangle_{ex}$ are 4.3 eV and 4.2 eV for Ce$_3$Cr$_8$ and La$_3$Cr$_8$ when $U$ = 3.0 eV, respectively. The change of $\triangle_{ex}$ is about one fifth of the change of $U$ values, e.g., $\triangle_{ex1}-\triangle_{ex2}=1/5(U_1-U_2)$; 3) Fit Eq~\ref{eq:forth} by setting $J_0$ obtained from step 1 as the y values, $U$ as the x values, and $V_{pd}^\pi$ as the fitting parameters. The corresponding parity plot with a unity slope is shown in Fig~\ref{fig_fit}(c) (see Supplementary materials Fig.~S3 for the fitted curves), which indicates the effective of Eq.~\ref{eq:forth} in describing the $U$ dependence of $J_0$. Overall, the superexchange model in the framework of t-e hybridization can correctly describe our DFT calculated results in Re$_3$Cr$_8$, confirming that the extraordinary FM dominated exchange coupling in the molecule is originated from the t-e hybridization, which is distinct with its bulk crystal due to the much larger distortion with even smaller TM-O-TM angle in the molecules. 

To understand the reason why $J_0$ in Ce$_3$Cr$_8$ (26.63 meV) is much small compared to that in La$_3$Cr$_8$ (52.22 meV), which is responsible to the much less sensitivity of $J_2$ with respect to the TM-O-TM angles in Ce$_3$Cr$_8$ (orange curve in Fig.~\ref{fig_fit}a) compared to that in La$_3$Cr$_8$ (green curve in Fig.~\ref{fig_fit}a), we can reasonably attribute to the obvious difference between Ce$_3$Cr$_8$ and La$_3$Cr$_8$, that at, the effect of the $f$ orbitals. The projected density of state (PDOS) plots (see Fig.~\ref{fig_dos}a, b) for FM state of both Ce$_3$Cr$_8$ and La$_3$Cr$_8$ show that Ce-$f$ orbitals are much closer to the Fermi level compared to the La-$f$. A higher hybridization between Ce-4f and O-2p is also observed in Ce$_3$Cr$_8$, e.g., the PDOS of Ce-4f orbitals and O-2p orbitals have similar peak in the same energy in Ce$_3$Cr$_8$ (see Fig.~\ref{fig_dos} c,d). The above evidence indicate that the VCT can happen in Ce-O but not in La-O. In addition, the fitted parameters $V_{pd}^\pi$ in Eq.~\ref{eq:forth} are 0.42 eV and 0.60 eV for Ce$_3$Cr$_8$ and La$_3$Cr$_8$, respectively. While the similar hopping terms calculated from WLWFs(see supplementary materials Table.~S5) and the same form of overlap integral in the Slater's approach\cite{slater1954simplified} indicate that the orbital overlap integral $V_{pd}^\pi$ should be close with each other in the two systems. This discrepancy also reminds us that a correction should be applied to the t-e hybridization framework to include the effects of $f$ orbitals.

\begin{figure*}[hbt]
\centering
\includegraphics[width=0.9\linewidth]{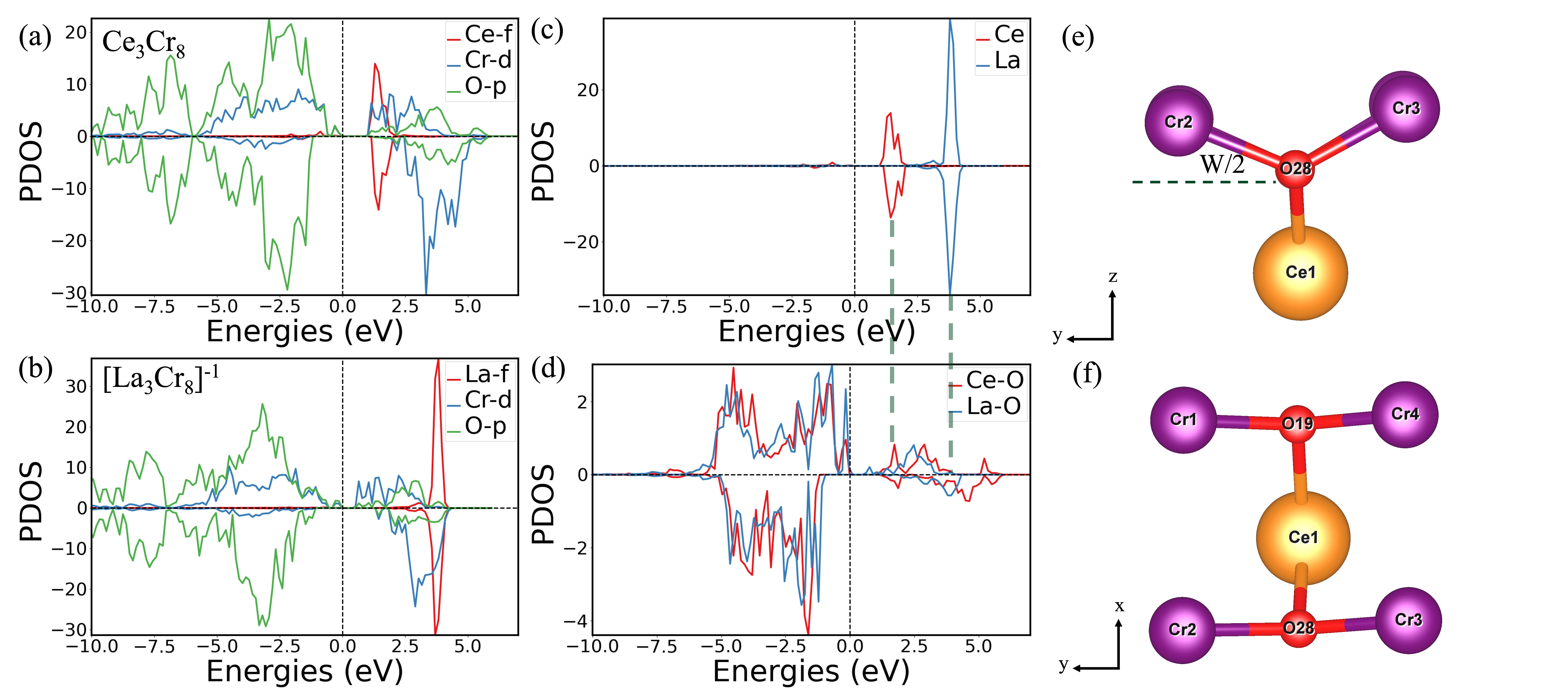}
\caption{The projected density of states (PDOS) for (a) FM(ground state) of Ce$_3$Fe$_8$ and (b) [La$_3$Fe$_8$]$^{-1}$ (c) Ce-f and La-f orbitals in Ce$_3$Cr$_8$ and La$_3$Cr$_8$ (d) O-p of J$_2$ in Ce$_3$Cr$_8$ and La$_3$Cr$_8$. The PDOS plots are corresponding orbital-resolved, respectively.(e,f) Schematic diagram of J$_2$ structure.}
\label{fig_dos}
\end{figure*}

%\subsection{\label{sec:effect}Effect of Ce-f orbital on t-e hybridization}

To include the $f$ orbitals in the process of VCT, we extend the t-e hybridization framework according to superexchange model\cite{goodenough1955theory,goodenough1958interpretation,kanamori1959superexchange,anderson1959new,slater1954simplified}(see supplementary materials section II for detailed analysis), as shown below,

\begin{equation} 
\label{eq:eight} 
\begin{aligned} 
J &= J_{d}+J_{f}
\end{aligned} 
\end{equation}
\begin{equation} 
\label{eq:nine} 
\begin{aligned} 
J_{d} &= \frac{(b_{pd}^\sigma)^2}{U+\triangle_c}-\frac{(b_{pd}^\pi)^2}{U+\triangle_{ex}}
\end{aligned} 
\end{equation}
\begin{equation} 
\label{eq:ten} 
\begin{aligned} 
J_{f} &=0.5[\frac{(b_{pd}^{\pi})^2}{U+\triangle_{ex}}-\frac{(b_{pd}^{\sigma})^2}{U+\triangle_c}]- \frac{(b_{pf}^\sigma)^2}{U+\triangle_f}+\frac{(b_{pf}^\pi)^2}{U+\triangle_f}
\end{aligned} 
\end{equation}
\begin{equation} 
\label{eq:eleven} 
\begin{aligned} 
b_{pd} &= c_{pd}\cdot V_{pd}; b_{pf} = c_{pf}\cdot V_{pf}
\end{aligned} 
\end{equation}
\begin{equation} 
\label{eq:twelve} 
\begin{aligned} 
J_{Ce} &=J_d+J_f =0.5[ \frac{(b_{pd\sigma}^{Ce})^2}{U+\triangle_c^{Ce}}-\frac{(b_{pd\pi}^{Ce})^2}{U+\triangle_{ex}^{Ce}}]
\end{aligned} 
\end{equation}
\begin{equation} 
\label{eq:thirdteen} 
\begin{aligned} 
J_{La} &=J_d = \frac{(b_{pd\sigma}^{La})^2}{U+\triangle_c^{La}}-\frac{(b_{pd\pi}^{La})^2}{U+\triangle_{ex}^{La}}
\end{aligned} 
\end{equation}
\begin{equation} 
\label{eq:fourteen} 
\begin{aligned} 
J_0^{La} &=\frac{(V_{pd\pi}^{La})^2}{U+\triangle_{ex}^{La}}; J_0^{Ce}=\frac{0.5(V_{pd\pi}^{Ce})^2}{U+\triangle_{ex}^{Ce}}
\end{aligned} 
\end{equation}
where J$_{d}$ and J$_{f}$ are the exchange coupling contribution from d orbitals and f orbitals, respectively.J$_{Ce}$ and J$_{La}$ are the exchange coupling contribution in Ce$_3$Cr$_8$ and La$_3$Cr$_8$ molecule, respectively. The different $b$ variables are transfer integrals of corresponding orbitals (see supplementary materials Section II for detailed formula). Eq.~\ref{eq:nine} is the same with Eq.~\ref{eq:third} with integrating the dimensionless $c$ parameters to the transfer integrals $b$, as shown in Eq. Eq.~\ref{eq:eleven}. For J$_f$ in Eq.~\ref{eq:ten}, the Ce-O bonds are approximately perpendicular to the J$_2$ path (the degree of $\angle$Cr-O-Ce is close to 90\textdegree) (see Fig~\ref{fig_dos} e-f), thus the symmetry in the Ce-O-Cr bonds are opposite to that in the Cr-O-Cr bonds, that is $\sigma$ and $\pi$ in Cr-O-Cr should be $\pi$ and $\sigma$ in Cr-O-Ce. To keep consistent notation of $\sigma$ and $\pi$, an opposite sign of $\frac{(b_{pd}^{\pi})^2}{U+\triangle ex}$ and $\frac{(b_{pd}^{\sigma})^2}{U+\triangle c}$ is assigned in J$_f$ in Eq.~\ref{eq:ten} compared to J$_d$ in Eq.~\ref{eq:nine}\cite{kanamori1959superexchange}. The Cr-O-Cr path has two O$^{2-}$-Cr$^{3+}$ exchange coupling interactions, while the Cr-O-Ce path only has one O$^{2-}$-Cr$^{3+}$ exchange coupling according to Anderson's mechanism\cite{anderson1959new}. This leads to the prefactor of pd superexchange coupling in J$_f$ is half of that in J$_d$. The $b_{pf}$ contributions from the VCT to unoccupied 4f orbitals in La$_3$Cr$_8$ can be ignored due to the negligible exchange interaction from the La-f orbitals, as discussed in the previous paragraph. However, even in Ce$_3$Cr$_8$, the b$_{pf}$ is still an order of magnitude smaller than b$_{pd}$ ( see supplementary materials Table.~S6). Therefore, the J$_0$ in Ce$_3$Cr$_8$ and La$_3$Cr$_8$ can be finally expressed in the form of Eq.~\ref{eq:fourteen}, which clearly indicates that $J_0$ in Ce$_3$Cr$_8$ (26.63 meV) is about half of that in La$_3$Cr$_8$ (52.22 meV) after considering the effects of $f$ orbitals. Noting that $\frac{(V_{pd\pi})^2}{U+\triangle_{ex}}$ in Ce$_3$Cr$_8$ and La$_3$Cr$_8$ are approximately the same.

\begin{table}[h]
\caption{The exchange coupling parameters ($J$) in meV of $\mathrm{Ce_3Cr_8}$ and $\mathrm{La_3Cr_8}$. The degree of the $\angle$Cr-O-Cr in J$_2$ is labeled. The difference of the parameters in $\mathrm{La_3Cr_8}$  with respect to those in $\mathrm{Ce_3Cr_8}$ are shown in parentheses.}
\label{table_J}
\begin{ruledtabular}
\begin{tabular}{ccc}
J Path   & Ce$_3$Cr$_8$  & La$_3$Cr$_8$  \\
\colrule

$J_1$(relaxed structure)& 1.09 & 1.54 (+0.45)  \\
$J_2$(relaxed structure)& 2.17 & 2.53 (+0.36) \\
125-$J_2$& 3.96 & 5.17 (+1.21) \\
127-$J_2$& 3.55 & 4.13 (+0.58)\\
129-$J_2$& 2.90 & 2.75 (-0.15)\\
131-$J_2$& 1.99 & 1.02 (-0.97)\\
133-$J_2$& 0.81 & -1.09 (-1.90)\\
\end{tabular}
\end{ruledtabular}
\end{table}

\textit{Discussion.--}We have theoretically investigated the atomic structure, electronic and magnetic properties of the molecule analogue of perovskite chromites, abbreviated as Ce$_3$Cr$_8$, using the first-principles method. In particular, we found a unexpected dominate FM interaction in magnetic coupling between two NN Cr sites with angle Cr-O-Cr much greater than 90\textdegree, distinct from the bulk perovskite chromites\cite{zhou2010intrinsic,zhu2022crystal}. The magnetic coupling between Cr ions evolving with the degree of $\angle$Cr-O-Cr follows the t-e hybridization framework, confirming that the dominated FM nature is originated from t-e hybridization. In particular, due to the great sensitivity of t-e hybridization with respect to the superexchange angle Cr-O-Cr,  little molecular structure change can induce significant difference in the magnetic exchange coupling strength with the ground state evolving a transition from FM to AF. This structure-dependent magnetic interaction mechanism can be applied to develop multifunctional molecule-based magnetic devices.

The effects of the Re-$f$ orbitals to the exchange interaction of TM ions was initially pointed out in Ce$_3$Mn$_8$ by contributing a direct FM interaction\cite{thuijs2017molecular}, which was also confirmed in Ce$_3$Fe$_8$\cite{wang2023molecular}. Our calculation results in Ce$_3$Cr$_8$ also indicate that the Ce-$f$ orbitals can place effects on the superexchange interaction between TM ions but with a different mechanism from the previous work. In Fig~\ref{fig_fit}c, we can find that Ce-$f$ orbitals can significantly change the sensitivity of $J_2$ with respect to the structure change, not just contributing FM or AF interaction. All the works on the analysis of the effects of $f$ orbitals show that only the orbitals near the Fermi level can have sizable contribution to the superexchange interaction. Hence the effects of $f$ orbitals on the t-e hybridization are closely related to the energy level position of the orbitals, which can be modulated by the type of cations and TM elements, as well as the molecular geometry. For example, charging the molecule can significantly change the molecular geometries and energy levels of the orbitals \cite{li2014conformational,wang2017cation}, thus in turn to modify the magnetic interactions within the molecules.

\textit{Conclusion.--}In conclusion, our investigation of the Ce$_3$Cr$_8$ magnetic molecule elaborates the complex magnetism involving both t-e hybridization and Re-$f$ orbitals at the nanoscale. The superexchange via the Cr-$d$ and O-$p$ orbitals can provide a dominate FM interaction for both $J_1$ and $J_2$ through strong t-e hybridization induced by the structure distortion. The superexchange interaction involving the t-e hybridization is very sensitive to the molecular structure, e.g. the ground state can undergo a FM to AFM transition upon the structural change, which can be further modulated by the Re-$f$ orbitals. In particular, the inclusion of Ce-$f$ orbitals can decrease the sensitivity of the change of the magnetic coupling with respect to the molecular structural changes. Our theoretical work complements the theory of superexchange magnetism involving t-e hybridization and adds another dimension to fine tune the magnetic properties of nanoscale molecules through Tm-d/Re-f interactions.

\begin{acknowledgments}
The authors would like to acknowledge financial support from the Hundreds of Talents Program of Sun Yat-sen University.
\end{acknowledgments}

\bibliographystyle{IEEEtran}
\bibliography{main}% Produces the bibliography via BibTeX.

\end{document}